\documentstyle[epsf]{lamuphys}
\begin{document}

\title{QUANTUM SPIN GLASSES}
\author{Heiko Rieger\inst{1} and A. Peter Young\inst{2}
}
\institute{
HLRZ c/o Forschungszentrum J\"ulich, 52425 J\"ulich, Germany
\and 
Department of Physics, University of California, Santa Cruz, CA 95064, USA
}
\maketitle

\begin{abstract}
Ising spin glasses in a transverse field exhibit a zero temperature
quantum phase transition, which is driven by quantum rather than
thermal fluctuations. They constitute a universality class that is
significantly different from the classical, thermal phase transitions.
Most interestingly close to the transition in finite dimensions a
quantum Griffiths phase leads to drastic consequences for various
physical quantities: for instance diverging magnetic susceptibilities
are observable over a whole range of transverse field values in the
disordered phase.
\end{abstract}

\newcommand{\be}{\begin{equation}}
\newcommand{\ee}{\end{equation}}

\section{Introduction}

At very low temperatures the role of quantum fluctuations in any
physical pure or disordered system become more and more important. As
far as critical phenomena are concerned any finite temperature
destroys quantum coherence of the lowest lying excitations determining
the universality class of the transition, which remains therefore
classical. However, if the transition takes place at strictly zero
temperature, a new universality class and in particular new physics
emerges. Obviously in the vicinity of such a transition even finite
temperature properties are characterized by strong crossover effects
between a quantum critical and classical regions. Thus the properties
of such zero temperature transitions become experimentally accessible,
which motivates the study these new universality classes.

The prominent feature of quantum phase transitions \cite{QPT} is the
fact that statics and dynamics are inextricably linked: the static
features are defined by the Hamiltonian, which implies immediately the
dynamics via the Schr\"odinger equation. This introduces an extra
dimension, the (imaginary) time, into the problem and it is by no means
guaranteed that this additional dimension is equivalent to one of the $d$
space dimensions.  In many {\it pure} systems it turns out to be so,
which is the origin of the observation that ``the correlation length
is proportional to the inverse energy gap''. This relation 
seems to be pretty robust even in the presence of a nontrivial
anisotropy (c.f.\ the ANNNI model), for which reason it became
folklore in strongly correlated systems.

However, in the presence of quenched (i.e.\ time-independent) {\it
disorder}, this simple relation fails: Usually the randomness
(modeling impurities etc.) is uncorrelated in space, but
time-independence means a perfect correlation of the disorder in the
time direction. This implies an {\it extreme} anisotropy, which
manifests itself in a non-trivial relation between spatial correlation
length $\xi$ and energy gap $\Delta E$. For a generic second order
phase transition scenario a dynamical exponent $z$ can be introduced
via $\xi\sim\Delta E^{-z}$, with $z$ in general different from one.

Another drastic consequence of the perfect correlation of the disorder
in the (imaginary) time direction are spectacular properties of physical
observables within the so called Griffiths phase \cite{griffiths}
surrounding the critical point itself. In contrast to the classical
case one there may be a whole region of values for the parameter
tuning the transition over which the zero-frequency susceptibilities
{\it diverge}. Such a scenario has actually been described already a
long time ago by McCoy \cite{mccoy} in the context of a one-dimensional
model. It seems that this is indeed a generic feature of finite
dimensional disordered systems with a quantum phase transition as 
we would like to demonstrate here.

In this article we focus on quantum spin glasses, in particular the
transverse field Ising models defined by the Hamiltonian
\be
H = -\sum_{\langle ij\rangle} J_{ij}\sigma_i^z\sigma_j^z
    -\Gamma\sum_i \sigma_i^x\;,
\label{hamilton}
\ee
where $\sigma_i$ are the Pauli spin matrices (modeling
spin-$\frac{1}{2}$ degrees of freedom), $\langle ij\rangle$ nearest
neighbor pairs on a $d$-dimensional lattice, $\Gamma$ the transverse
field and $J_{ij}$ quenched random interaction strength obeying for
instance a Gaussian distribution with zero mean and variance one.
Here the quenched disorder mentioned above also produces frustration,
which enhances the effect that the randomness has on critical and
non-critical properties. For completeness we also include a discussion
of the one-dimensional case, which is not frustrated but shares many
features of the higher dimensional realizations. In the next section
we report on the results for the critical properties and in section 3
we present the exciting features of the Griffiths phase in these
models. Section 4 gives a summary and perspectives for future work.

\section{Critical Properties}

In any dimension $d$ the quantum system described by eq.\ 
(\ref{hamilton}) has a second order phase transition at zero
temperature that manifests itself in specific macroscopic properties
of the ground state and low lying excitations. A critical value
$\Gamma_c$ for the transverse field strength separates a disordered or
paramagnetic phase for $\Gamma>\Gamma_c$ from an ordered phase for
$\Gamma<\Gamma_c$. This transition is characterized by a diverging
length scale $\xi\sim|\Gamma-\Gamma_c|^{-\nu}$ and a vanishing
characteristic frequency $\omega\sim\Delta E\sim\xi^{-z}$. The latter
is the quantum analog of ``critical slowing down'' in the critical
dynamics of classical, thermally driven transitions. Together with a
third critical exponent, defining the anomalous dimension of the order
parameter field, the thermal exponent $\nu$ and the dynamical exponent
$z$ give a complete description of the transition via a set of
scaling relations for two and three dimensions. The one-dimensional
case shows a somewhat richer scenario \cite{fisher} and in $d>8$ the
violation of hyperscaling adds another exponent \cite{read}.  
We list the main features for the different dimensions:

\subsection{d=1}

Analytical \cite{mccoy,fisher} and numerical \cite{crisanti,young}
investigations revealed the following picture: because of the
logarithmically broad distribution of various physical quantities at
criticality one is forced discriminate between average and typical
properties. For instance the typical correlation length diverges with
an exponent $\nu_{\rm typ}=1$, whereas the average diverges with
$\nu_{\rm av}=2$. It turns out that it is not the energy gap but its
logarithm that scales with system size and distance from the critical
point, giving rise to an exponential rather than algebraic decrease of
the energy gap with system size: $[\Delta E]_{\rm
  av}\sim\exp(-aL^{1/2})$. This means that $z=\infty$ and since the
inverse gap corresponds to a characteristic relaxation or tunneling
time it is reminiscent of an activated dynamics scenario in
conventional spin glasses or random field systems.

For numerical work with finite size systems it is most useful to study
the probability distribution of various quantities, such as the energy
gap or local susceptibility $\chi^{({\rm loc})}(\omega=0)$. One finds
at the critical point
\be
P_L ( \ln \chi^{({\rm loc})} ) \sim
\tilde{P}(L^{-1/2}\,\ln \chi^{\rm loc})
\label{p_scale}
\ee
reflecting both features $\nu_{\rm av}=2$ (since the system size
enters with a power $1/\nu_{\rm av}$) and $z=\infty$ (since $\ln
\chi^{\rm (loc)}$ rather than $\chi^{\rm (loc)}$ enters the scaling
variable).

\subsection{d = 2 and 3}

Numerical investigation of the two- \cite{rieger} and
three-dimensional \cite{guo} quantum Ising spin glass model in a
transverse field via quantum Monte Carlo simulations 
demonstrated the existence of a second order phase transition at a
critical transverse field strength $\Gamma_c$. The results are
compatible with the scaling predictions made by a droplet theory for
these models \cite{thill}.

In contrast to the one-dimensional case there is ample evidence that 
here the dynamical exponent $z$ is finite, namely $z=1.50\pm0.05$ in
two and $z=1.3\pm0.1$ in three dimensions. Also the subsequent study
\cite{rieger2,guo2} of the probability distribution of the logarithms
of the local linear and nonlinear susceptibility at criticality gave
no indication of a characteristic broadening with increasing system
size that would indicate an infinite value for $z$. 

Most interestingly, as mentioned in the introduction, the critical
exponents for the quantum phase transition at strictly zero
temperature are connected to the temperature dependence of various
quantities for $\Gamma=\Gamma_c$, $T\to0$. For instance 
the experimentally accessible nonlinear susceptibility {\it diverges}
at $\Gamma=\Gamma_c$  quite strongly, i.e.\
\be
\chi_{\rm nl}(\omega=0) 
= \frac{\partial^3 M}{\partial h^3}\bigg\vert_{h=0}
\propto T^{-\gamma_{\rm nl}/z}
\ee
with $\gamma_{\rm nl}/z\approx3.0$ in two and in three dimensions ($M$
is the total magnetization and $h$ a longitudinal magnetic field).
Thus the strength of the divergence of the non-linear susceptibility
is similar to the one at the classical transition. This is in sharp
contrast to the results reported for recent experiments on
LiHo$_x$Y$_{1-x}$F$_4$ \cite{exp}. Here the strength of the divergence
seemed to decrease with decreasing temperature, and it has been
speculated that at zero temperature, i.e.\ at the quantum phase
transition, no divergence at all might be present (see next
subsection). On the other hand, it might be possible that because of
the long range nature of the dipolar interactions in
LiHo$_x$Y$_{1-x}$F$_4$the experiments might be better described by 
a model with appropriately chosen long range interactions.

\subsection{Mean field (d=$\infty$)}

The quantum phase transition occuring in the infinite range model of
the transverse Ising spin glass \cite{miller} and in a related quantum
rotor spin glass \cite{ye,read} can be handled analytically to a large
extent \cite{remark}. The dynamical exponent is $z=2$, the correlation
length exponent $\nu=1/4$ and the nonlinear susceptibility diverges
with an exponent $\gamma=1/2$. The latter again implies a divergence
of the non-linear susceptibility. Although it is much weaker than for
short range interaction models it should clearly be
observable. However, the experiments \cite{exp} at small but finite
temperatures yield an effective exponent $\gamma_{\rm eff}$ that is
significantly smaller than $1/2$, and for $T<25\,{\rm mK}$ even one
that is indistinguishable from zero. Therefore also mean-field theory
does not cure the contradiction between experiment and theory for
transverse Ising spin glasses.

Finally we would like to remark that the zero temperature phase transition in
a number of mean field quantum spin glasses falling in universality
classes different from the one considered here has been investigated
recently \cite{other_qsg}.

\section{Griffiths Phase}

In a specific disorder realization one can always identify spatial
regions that are more weakly or more strongly coupled than the
average. The latter lead to a non-analytic behavior of the free energy
not only at the critical point but in a whole region surrounding it,
as already observed by Griffiths nearly 30 years ago \cite{griffiths}.
However, a strongly coupled cluster (in spin glass terms one with a
minor degree of frustration) tends to order locally much earlier,
coming from the disordered phase, than the rest of the system. Upon
application of an external field spins in this cluster act
collectively and thus lead locally to a greatly enhanced
susceptibility. However, one only gets an essential singularity in
static properties of a classical system \cite{griffiths}.

This effect is much stronger in dynamics than in statics.
For classical disordered Ising system with heat bath
dynamics close to a thermally driven phase transition it was shown
\cite{randeira} that these strongly coupled regions posses very large
relaxation times for spin autocorrelations. By averaging over all
sites one gets a so called enhanced power law
\be
C_{\rm classical}(t)=[\langle S_i(t)S_i(0) \rangle]_{\rm av}
  \sim\exp\Bigl\{-a(\ln t)^{d/d-1}\Bigr\}
\label{ct_class}
\ee
where $[\ldots]_{\rm av}$ means an average over all sites and disorder
realizations. However, this prediction, although rigorous to some
extent (at least for the diluted ferromagnet \cite{rigorous}), could
not be confirmed by extensive numerical investigations, neither for
the spin glass \cite{sg_mc} nor for the diluted ferromagnet
\cite{dil_mc,dil_exp}. The main reason for this failure is most probably the
extremely small probability with which big enough strongly coupled
clusters that have an extremely large relaxation time occur in a
finite size systems \cite{sg_prob}.

To make this point clear let us focus, for simplicity, on a diluted
Ising ferromagnet below the percolation threshold, i.e.\ the site
concentration $c<p_c$. Then a connected cluster of volume $V=L^d$
occurs with a probability $p=c^V=\exp(-\zeta L^d)$ with $\zeta=\ln
1/c > 0$, which is a very small number for large $L$. However, these
spins within this cluster have a relaxation time $\tau$ that is
exponentially large (because of the activation energy that is needed
to pull a domain wall through the cluster to turn it over):
$\tau\sim\exp(\sigma L^{d-1})$, with $\sigma$ a surface tension. the
combination of these two exponentials yield the result (\ref{ct_class}).
Note that the occurrence of the exponent $d-1$ in the relaxation time
renders the final decay of the autocorrelation function {\it faster}
than algebraic.

This is quite different in a random quantum system at zero
temperature. Now the activated dynamics in the classical scenario has
to be replaced by tunneling events. Let us stick to the above example
of a diluted ferromagnet and add a transverse field at zero
temperature. The probability for a connected cluster to occur is the
same as before, but the relaxation time or inverse tunneling rate is
now given by $\tau\sim\exp(\sigma'L^d)$.  Remember that the classical
ground state of the cluster under consideration would be ferromagnetic
and twofold degenerate. The transverse field lifts this degeneracy for
a finte cluster, but for $\Gamma$ smaller than the critical field
value, where the ferromagnetic order of the infinite pure system sets
in, the energy gap is extremely small: it decreases exponentially with
the volume of the cluster. This energy gap between ground state and
first excited state sets the scale for the tunneling rate.

Now we see that in marked contrast to the classical system a $d$
instead a $d-1$ appears in the exponent for the relaxation time.  One
might say that the same cluster relaxes faster via activated dynamics
(although that is already incredibly slow) than by quantum tunneling.
Therefore, by the same line of arguments that lead to
(\ref{ct_class}), one now expects spin-autocorrelations to decay {\it
algebraically} in the Griffiths phase of random quantum systems:
\be
C_{\rm quantum}(t)=\langle0\vert\sigma_i^z(t)\sigma_i^z(0)\vert0\rangle
    \sim t^{-\zeta(\Gamma)/\sigma'(\Gamma)}
\label{ct}
\ee
For simplicity, although we expect a similar form for real time
correlations, we assume $t$ to be imaginary time, i.e.\ the Operator
$\sigma_i^z(t)$ to be given by $\exp(-Ht)\,\sigma_i^z\,\exp(Ht)$, $H$
being the Hamiltonian. Apart from a different functional form
(\ref{ct}) has drastic consequences for various quantities: For
instance the local zero frequency susceptibility at temperature $T$ is
given by
\be
\chi^{({\rm loc})}(\omega=0)=\int_0^{1/T}dt\,\langle C(t)\rangle
\sim T^{-1+\zeta(\Gamma)/\sigma'(\Gamma)}
\label{temp}
\ee
which {\it diverges} for $T\to0$ if
$\zeta(\Gamma)/\sigma'(\Gamma)<1$ even if one is not at the critical point!
Such a prediction should be experimentally measurable.

It should be noted that a diverging (surface) susceptibility in a
whole region close to the critical point has already been found by
McCoy nearly 30 years ago \cite{mccoy} for a classical two-dimensional
Ising model with layered randomness, which is equivalent to the random
transverse Ising chain. Actually we see now clearly that the
non-analyticities reported by Griffiths \cite{griffiths} and the
divergences calculated by McCoy \cite{mccoy} have a common origin: the
existence of strongly coupled rare regions, which simply leads to a
more drastic effect in a quantum system at low or vanishing
temperatures than in a classical system.

\subsection{d = 1}

The random transverse Ising chain has been reinvestigated recently by
D.\ Fisher \cite{fisher} in a beautiful renormalization group
treatment providing us with many astonishing analytical
predictions. On the numerical side one can take advantage of the free
fermion technique \cite{lsm}, by which the problem of diagonalizing
the original quantum Hamiltonian is reduced to the diagonalization of
a $L\times L$ or $2L\times 2L$ matrix. In this way one can consider
very large system sizes ($L\sim256$) with good statistics (i.e.\
number of disorder realizations $\ge50,000$). However, we should note
that some of the basic features of the model (e.g.\ $z=\infty$ at
criticality, existence of a Griffiths phase, etc.) can already be seen
for modest system sizes ($L\le16$) provided one studies the
appropriate quantities which are sensible to the rare disorder
configurations having strong couplings and/or small transverse fields.
The probability distribution of the energy gap and/or
various susceptibilities is such a quantity and in what follows we
report the results of numerical investigations that support the above
predictions.

In \cite{young} we took a uniform distribution for bonds
($J_i\in[0,1]$) and transverse fields ($h_i\in[0,h_0]$). The system is
at its critical point for $h_0=1$ (since in this case the field- and
bond distributions are identical and therefore the model self-dual)
and in the paramagnetic (ferromagnetic) phase for $h_0>1$
($h_0<1$). For each of 50,000 disorder realization for chains of
length $L$ (from $L=16$ to $L=128$) we calculated the energy gap
$\Delta$ from which we calculated the probability distribution
$P_L(\ln\Delta E)$. It turned out that {\it at} the critical point
$h_0=1$ the probability distribution scales as described in section
2.1.\ according to eq.\ (\ref{p_scale}). In particular the
distribution gets broader even on a logarithmic scale with increasing
system size.

Within the Griffiths phase for $h_0>1$, however, the distribution does not
broaden any more, but gets simply shifted on a logarithmic scale with
increasing system size. It turns out that now the distribution of gaps
obeys
\be
\ln \left[ P(\ln \Delta E) \right] = {1\over z} \ln \Delta E + \mbox{const.} 
\label{p_griff}
\ee 
with a dynamical exponent that varies continuously throughout the
Griffiths phase according to $1/z\sim2\delta-c\delta^2$ ($\delta$
being the distance from the critical point $\delta=h-h_0$) and which
diverges at the critical point $z\to\infty$ for $\delta\to0$.
The form (\ref{p_griff}) can be made plausible by an argument applicable 
to any dimension $d$ given in the next subsection.

\subsection{d = 2 and 3}

For the numerical investigation of the astonishing features of the
transverse field spin glass model defined in eq.\ (\ref{hamilton}) in
2 and 3 dimensions it is also most promising to focus on the
probability distribution of the local linear and nonlinear
susceptibility. The former is simply proportional to the inverse of the
energy gap, whereas the latter is proportional to the third power of
the inverse of the gap, hence the distribution of gaps $P(\ln\Delta E)$ plays again
the crucial role. The form that we expect is similar to
(\ref{p_griff}) because of the following argument:

From what has been said in the beginning of section 3 before eq.\
(\ref{ct}), by combining the probability of a strongly coupled cluster
with its energy gap one gets a power law for the tail of the
distribution of gaps: $P(\Delta E)\sim\Delta E^{\lambda-1}$ with
$\lambda=\zeta(\Gamma)/\sigma'(\Gamma)$. Furthermore since the
excitations that give rise to a small energy gap are well localized we
assume that their probability is proportional to the spatial volume
$L^d$ of the system. Thus $P_L(\ln\Delta E)\sim L^d\Delta E^\lambda$,
which can be cast into the conventional scaling form relating a time
scale to a length scale via the introduction of a dynamical exponent:
$P_L(\ln\Delta E)\sim (L\Delta E^{1/z})^d$ with $z=d/\lambda$. Therefore
the distribution for the local susceptibility (because $\chi^{\rm
(loc)} \propto 1/\Delta E$) should be given by
\begin{equation}
\label{tail_chi}
\ln \left[ P(\ln \chi^{\rm (loc)} ) \right] =
-{d\over z} \ln \chi^{\rm (loc)} + \mbox{const.}
\end{equation}
As a byproduct one obtains in this way also the form (\ref{p_griff})
for the one-dimensional case.  The distribution of the local nonlinear
susceptibility the factor $d/z$, which is actually the power for the
algebraically detaying tail of $P(\chi^{\rm (loc)})$, should be
replaced by $d/3z$ (since $\chi^{\rm (loc)}_{\rm nl}\propto1/\Delta
E^3$):
\begin{equation}
\ln \left[ P(\ln \chi_{\rm nl}^{\rm (loc)} ) \right] =
-{d\over 3z} \ln \chi_{\rm nl}^{\rm (loc)} + \mbox{const.}
\label{tail_chi_nl}
\end{equation}
Thus one can characterize {\it all} the Griffiths singularities in the
disordered phase by a {\it single} continuously varying exponent, $z$.
This prediction is indeed confirmed by the numerical investigations
\cite{rieger2,guo2} as demonstrated in fig.\ 1.
\vskip-0.5cm
\begin{figure}
\epsfxsize=12cm\epsfbox{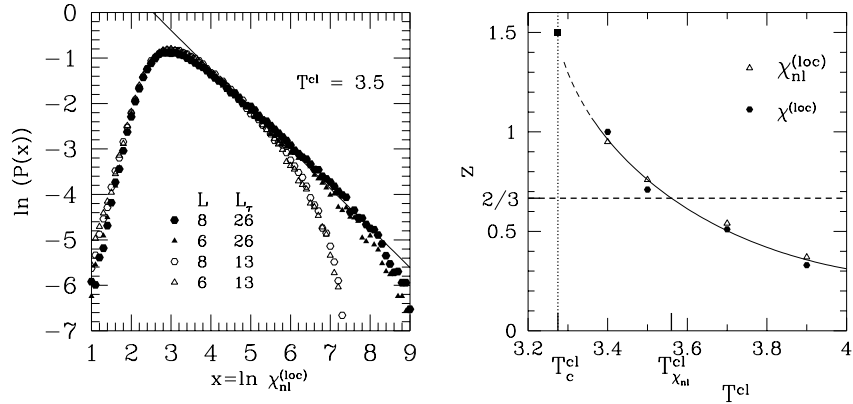}
\vskip-0.5cm
\caption{\footnotesize\baselineskip=8pt
  \label{fig1}
  {\bf Left:} The probability distribution of the local non-linear
  susceptibility at a point in the disordered phase (the critical
  point is at $T_c^{\rm cl}=3.30)$. The parameter $T^{\rm cl}$ is
  related to the transverse field strength via the Suzuki-Trotter
  mapping (see [9] for details). 
  The straight line has slope $-0.87$ 
  which gives via (\protect{\ref{tail_chi_nl}}) $z= 0.76$.
. Since the slope is greater than $-1$, or equivalently $ z > 2/3$, 
  the average non-linear susceptibility does {\it diverge} at this
  point. {\bf Right:} The 
  dynamical exponent $z$, obtained by fitting the distributions of
  $\chi^{\rm (loc)}$ and $\chi^{\rm (loc)}_{\rm nl}$ to
  Eqs.~(\protect\ref{tail_chi}) and (\protect\ref{tail_chi_nl}), is
  plotted for different values of $T^{\rm cl}$. The estimates obtained
  from data for $\chi^{\rm (loc)}$ are shown by the triangles and the
  estimates from the data for $\chi^{\rm (loc)}_{\rm nl}$ are shown by
  the hexagons. The two are in good agreement. The dotted vertical
  line indicates the critical point, obtained in
  Ref.~\protect\cite{rieger} and the solid square indicates the
  estimate of $z$ at the critical point.  The dashed line is $z=2/3$;
  the average non-linear susceptibility diverges for $z$ larger than
  this, i.e. $T^{\rm cl} > T^{\rm cl}_{\chi_{\rm nl}} \simeq 3.56$.
  The solid curve is just a guide to the eye.}
\end{figure}
\vskip-0.2cm

We see that the {\it average} susceptibility in the disordered phase
($\Gamma>\Gamma_c$) will diverge when $z>d$, and the nonlinear
susceptibility when $z>d/3$. From fig.\ 1 (taken from \cite{rieger2})
we see that in two dimensions this is indeed the case for the
non-linear susceptibility in a range
$\Gamma\in[\Gamma_c,\Gamma_{\chi_{\rm nl}}]$. However, the numerical
value of $\Gamma_c$ and $\Gamma_{\chi_{\rm nl}}$ might be
non-universal and so we cannot make a precise prediction for what the
range of field-values, in which $\chi_{\rm nl}$ diverges, might be in
an experiment.

In three dimensions \cite{guo2} the situation is similar, although
here the range of transverse field values where the nonlinear
susceptibility diverges is much smaller than in 2d. This is plausible
since for increasing dimension the spectacular effects of the
Griffiths singularities, which have their origin in spatially
isolated clusters, becomes weaker and seem to vanish in the mean field
theory.

Finally we would like to emphasize that due to the divergent behavior
of the linear or higher susceptibilities in part of the quantum
Griffiths phase we expect this effect to be observable also in an
experiment with a macroscopic but finite system. It has been argued by
Imry \cite{imry} that the Griffiths singularity occuring in a {\it
  classical} diluted Ising magnet is an artifact of the thermodydnamic
limit procedure. What is meant here is the essential singularity in a
{\it static} quantity (like the magnetization or susceptibility).  In
essence the argument is that in order for these effects to be
observable one needs a macroscopic cluster, which would occur only
once in an astronomically large collections of finite samples.
However, since in the {\it quantum} case, which we consider here, the
statics is inextricably mixed with the {\it dynamics}, static
quantities are much more sensible with respect to the existence of
strongly coupled regions even of modest size. The distribution of
cluster sizes $n$ is still effectively cut off around $n_{\rm max}\sim
\log N$, where $N=L^d$ is the system size. However, such a {\it
  typical} cluster, which occurs with probability ${\cal O}(1)$ leads
to a (local) nonlinear susceptibility of $\chi_{nl}^{\rm (loc)}\sim
N^{3\sigma'/\zeta}$. For a macroscopic sample ($N\sim10^{23}$) this is
a huge number, which should render the divergence predicted for an
infinite sample observable also in a macroscopic {\it typical} system
(in contrast to the mere non-analyticities in the classical case).

\section{Conclusions}

The quantum phase transition occuring at zero temperature in transverse
field Ising spin glasses can be described by a set three independent
exponents that have been determined numerically. The dynamical
exponent $z$, connecting time- or energy scales with a spatial
diverging length is finite in $d>1$ and infinite in the
one-dimensional model. The critical point is surrounded by a quantum
Griffiths phase, where various susceptibilities diverge. These
divergences can be characterized by a single continuously varying
exponent $z$. Its limiting value seems to be identical to the critical
value, which is analytically established in 1d. The most important
observation is that, although a Griffiths phase should be present in
all random magnets, this is the first numerical verification of some
of its theoretically predicted features. We would not be surprised if
the divergences we found in the quantum Griffiths phase could also be
measured in an appropriate experiment. It is interesting to speculate
on whether the possible divergence of the non-linear susceptibility in
part of the disordered phase might be related to the difference
between the experiments \cite{exp} which apparently do not find a
strong divergence of $\chi_{\rm nl}$ at the critical point, and the
simulations\cite{rieger,guo} which do. Certainly more work in this
direction has to be done.


\begin{thebibliography}

\bibitem{[1]}{QPT}{[1]}
        S.\ Sachev, Z.\ Phys.\ B {\bf 94}, 469 (1994); Nucl. Phys. B
        {\bf 464}, 576 (1996); STATPHYS 19, ed.\ Hao Bailin, p.\ 289
        (World Scientific, Singapore, 1996).

\bibitem{[2]}{griffiths}{[2]}
        R. B. Griffiths, Phys. Rev. Lett. {\bf 23}, 17 (1969).

\bibitem{[3]}{mccoy}{[3]}
        B. McCoy, Phys. Rev. Lett. {\bf23}, 383 (1969).

\bibitem{[4]}{fisher}{[4]}
        D. S. Fisher, Phys. Rev. Lett. {\bf 69}, 534 (1992); 
        Phys. Rev. B {\bf 51}, 6411 (1995).

\bibitem{[5]}{read}{[5]}
        N. Read, S. Sachdev and J. Ye, 
        Phys. Rev. B {\bf 52}, 384 (1995).

\bibitem{[6]}{crisanti}{[6]}
        A. Crisanti and H. Rieger, 
        J. Stat. Phys. {\bf 77}, 1087 (1994).

\bibitem{[7]}{young}{[7]}
        A. P. Young and H. Rieger,
        Phys. Rev. B {\bf 53}, 8486 (1996).      

\bibitem{[8]}{rieger}{[8]}
        H. Rieger and A. P. Young, 
        Phys. Rev. Lett. {\bf 72}, 4141 (1994).
        
\bibitem{[9]}{guo}{[9]}
        M. Guo, R. N. Bhatt and D. A. Huse, 
        Phys. Rev. Lett. {\bf 72}, 4137 (1994).

\bibitem{[10]}{thill}{[10]}
        M. J. Thill and D. A. Huse, 
        Physica A, {\bf 15}, 321 (1995).

\bibitem{[11}{rieger2}{[11}
        H. Rieger and A. P. Young, 
        Phys. Rev. B {\bf 54}, *** (1996).      

\bibitem{[12]}{guo2}{[12]}
        M.~Guo, R.~N.~Bhatt and D.~A.~Huse,
        Phys. Rev. B {\bf 54}, *** (1996).

\bibitem{[13]}{exp}{[13]} 
        W. Wu, B. Ellmann, T. F. Rosenbaum, G. Aeppli and D. H. Reich,
        Phys. Rev. Lett. {\bf 67}, 2076 (1991); 
        W. Wu, D. Bitko, T. F. Rosenbaum and G. Aeppli,
        Phys. Rev. Lett. {\bf 71}, 1919 (1993).

\bibitem{[14]}{miller}{[14]}
        J. Miller and D. Huse, Phys. Rev. Lett. {\bf 70}, 3147 (1993).

\bibitem{[15]}{ye}{[15]}
        J. Ye, S. Sachdev and N. Read, 
        Phys. Rev. Lett. {\bf 70}, 4011 (1993).

\bibitem{[16]}{remark}{[16]} The finite temperature features of this model
        seem to be slightly more delicate beacause of the issue of
        replica symmetry breaking.  See Y. Y. Goldschmidt and
        P. Y. Lai, Phys. Rev. Lett. {\bf 64}, 2567 (1990) and B.\ K.\
        Chakrabarti, A.\ Dutta and P.\ Sen: {\it Quantum Ising Phases
        and Transitions in Transverse Ising Models}, Lecture Notes in
        Physics {\bf m41}, Springer Verlag,
        Berlin-Heidelberg-New$\,$York, (1996) for references.

\bibitem{[17]}{other_qsg}{[17]}
        R.\ Oppermann and M.\ Binderberger, 
        Ann. Physik {\bf 3}, 494 (1994);
        S. Sachdev, N. Read and R.\ Oppermann, 
        Phys.~Rev.~B {\bf 52}, 10286 (1995);
        see also R.\ Oppermann in this volume.
        T. M. Nieuwenhuizen, 
        Phys.\ Rev.\ Lett.\ {\bf 74}, 4289 (1995).
        F. P\'azm\'andi, G. T. Zim\'anyi and R. T. Scalettar,
        cond-mat/9602158 (1996).

\bibitem{[18]}{randeira}{[18]}
        M.\ Randeira, J.\ P.\ Sethna and R.\ G.\ Palmer,
        Phys.~Rev.~Lett.\ {\bf 54}, 1321 (1985).

\bibitem{[19]}{rigorous}{[19]}
        A.~J.~Bray, Phys.~Rev.~Lett. {\bf 60}, 720 (1988), 
        J.~Phys.~A {\bf 22}, L81 (1989);
        D.~Dhar, M.~Randeira and J.~P.~Sethna,
        Europhys.~Lett. {\bf 5}, 485 (1988).
        
\bibitem{[20]}{sg_mc}{[20]}
        A.~T.~Ogielski, Phys.~Rev.~B {\bf 32}, 7384 (1985);\\
        H.~Takano and S.~Miyashita,
        J.~Phys.~Soc.~Jap. {\bf 64}, 423 (1995).

\bibitem{[21]}{dil_mc}{[21]} 
  S.~Jain,~J.~Phys.~C~{\bf 21},~L1045~(1988);\hfill\break
        H.~Takano, S.~Miyashita, 
        J.~Phys.~Soc.~Jap. {\bf 58}, 3871 (1989);\hfill\break
        V.~B.~Andreichenko, W.~Selke and A.~L.~Talapov,
        J.~Phys.~A {\bf 25}, L283 (1992);\hfill\break
        P.~Grassberger, preprint (1996).

\bibitem{[22]}{dil_exp}{[22]}
        Experimentally one finds qualitative indications
        for the existence of a classical Griffiths phase e.g.\ in diluted
        (anti)ferromagnets see: Ch. Binek and W. Kleemann, Phys.\
        Rev.\ B {\bf 51}, 12888 (1995).

\bibitem{[22]}{sg_prob}{[22]}
        For an investigation of the probability distribution of
        local relaxation times in classical spin glasses within the
        Griffiths phase see: H.\ Takayama, T.\ Komori and K.\
        Hukushima, in {\it Coherent Approach to Fluctuations}, ed.\ M.\
        Suzuki and N.\ Kawashima, p.\ 155 (World Scientific, Singapore, 1996);
        {\it ibid.} H.\ Takano and S.\ Miyashita, p.\ 217.

\bibitem{[23]}{lsm}{[23]}
        E.~Lieb, T. Schultz and D.~Mattis, 
        Ann. Phys. (NY) {\bf 16}, 407 (1961).

\bibitem{[24]}{imry}{[24]}
        Y.\ Imry, Phys.\ Rev.\ B {\bf 15}, 4448 (1977).

\end{thebibliography}
\end{document}